\documentclass[aps,prd,superscriptaddress,showpacs,preprint]{revtex4}
\usepackage{graphicx, bm}
\usepackage[usenames]{color}

\begin{document}
\draft
\title{Bounds on the dipole moments of the tau-neutrino via the process $e^{+}e^{-}\rightarrow \nu \bar
\nu \gamma$ in a 331 model}

\author{ A. Guti\'errez-Rodr\'{\i}guez}
\affiliation{\small Facultad de F\'{\i}sica, Universidad Aut\'onoma de Zacatecas\\
         Apartado Postal C-580, 98060 Zacatecas, M\'exico.\\}

\date{\today}

\begin{abstract}

We obtain limits on the anomalous magnetic and electric dipole
moments of the $\nu_{\tau}$ through the reaction
$e^{+}e^{-}\rightarrow \nu \bar \nu \gamma$ and in the framework
of a 331 model. We consider initial-state radiation, and neglect
$W$ and photon exchange diagrams. The results are based on the
data reported by the L3 Collaboration at LEP, and compare
favorably with the limits obtained in other models, complementing
previous studies on the dipole moments.
\end{abstract}

\pacs{14.60.St, 13.40.Em, 12.15.Mm, 12.60.-i
\vspace*{2cm}\noindent  E-mail: $^{1}$alexgu@fisica.uaz.edu.mx}

\vspace{5mm}

\maketitle

\section{Introduction}
In the Standard Model (SM) \cite{S.L.Glashow,Weinberg,Salam} extended to contain
right-handed neutrinos, the neutrino magnetic moment induced by
radiative corrections is unobservably small, $\mu_\nu \sim 3\times
10^{-19}(m_\nu/1 \hspace{1mm} eV)$ \cite{Mohapatra}. Current
limits on these magnetic moments are several orders of magnitude
larger, so that a magnetic moment close to these limits would
indicate a window for probing effects induced by new physics
beyond the SM \cite{Fukugita}. Similarly, a neutrino electric
dipole moment will point also to new physics and they will be of
relevance in astrophysics and cosmology, as well as terrestrial
neutrino experiments \cite{Cisneros}.

The existence of a heavy neutral ($Z'$) vector boson is a feature
of many extensions of the standard model. In particular, one (or
more) additional $U(1)'$ gauge factor provides one of the simplest
extensions of the SM. Additional $Z'$ gauge bosons appear in Grand
Unified Theories (GUTs) \cite{Robinett}, Superstring Theories
\cite{Green}, Left-Right Symmetric Models (LRSM)
\cite{Mohapatra,G.Senjanovic,G.Senjanovic1} and in other models
such as models of composite gauge bosons \cite{Baur}. In
particular, it is possible to study some phenomenological features
associates with this extra neutral gauge boson through models with
gauge symmetry $SU(3)_C\times SU(3)_L\times U(1)_N$, also called
331 models. These models arise as an interesting alternative to
explain the origin of generations. Pisano and Pleitez
\cite{Pisano} have proposed an model based on the gauge group
$SU(3)_C\times SU(3)_L\times U(1)_N$. This model has the
interesting feature that each generation of fermions is anomalous,
but that with three generations the anomalous canceled. Detailed
discussions on 331 models can be found in the literature
\cite{Pisano,Montero,Hoang}.

T. M. Gould and I. Z. Rothstein \cite{T.M.Gould} reported in 1994
a bound on $\mu_{\nu_\tau}$ obtained through the analysis of the
process $e^{+}e^{-} \rightarrow \nu \bar \nu \gamma$, near the
$Z_1$-resonance, with a massive neutrino and the SM
$Z_1e^{+}e^{-}$ and $Z_1\nu \bar \nu$  couplings.

At low center of mass energy $s << M^2_{Z_1}$, the dominant
contribution to the process $e^+e^- \to \nu \bar\nu \gamma$
involves the exchange of a virtual photon \cite{H.Grotch}. The
dependence on the magnetic moment comes from a direct coupling to
the virtual photon, and the observed photon is a result of 
initial-state Bremsstrahlung.

At higher s, near the $Z_1$ pole $s\approx M^2_{Z_ 1}$, the
dominant contribution involves the exchange of a $Z_1$ boson. The
dependence on the magnetic moment $(\mu_{\nu_\tau})$ and the
electric dipole moment $(d_{\nu_\tau})$ now comes from the
radiation of the photon observed by the neutrino or antineutrino
in the final state. We emphasize here the importance of the final
state radiation near the $Z_1$ pole of a very energetic photon as
compared to conventional Bremsstrahlung.

However, in order to improve the limits on the magnetic moment and
the electric dipole moment of the tau-neutrino, in our calculation
to the process $e^+e^- \to \nu \bar\nu \gamma$ we consider 
initial-state radiation, in this way the bounds on the dipole moments are
stronger than those evaluated in previous studies by other
authors. We neglect $W$ and photon exchange diagrams, which amount
to $1\hspace{.5mm} \%$ corrections in the relevant kinematic
regime. The Feynman diagrams which give the most important
contribution to the cross section are shown in Fig. 1.

Our aim in the present paper is to analyze the reaction
$e^{+}e^{-}\rightarrow \nu \bar\nu \gamma$ in the framework of a
331 model and we attribute an anomalous magnetic moment (MM) and
an electric dipole moment (EDM) to a massive tau-neutrino. This
process serve to set limits on the tau-neutrino MM and EDM. In
this paper, we take advantage of this fact to set limits on
$\mu_{\nu_{\tau}}$ and $d_{\nu_{\tau}}$ for various values of the
mixing angle $\phi$ of the 331 model, according to Refs.
\cite{Cogollo, Hoang}.

The L3 Collaboration \cite{L3} evaluated the selection efficiency using
detector-simulated $e^{+}e^{-}\rightarrow \nu \bar \nu \gamma
(\gamma)$ events, random trigger events, and large-angle
$e^{+}e^{-}\rightarrow e^+e^-$ events. From Fig. 1 of Ref. \cite{L3} the process
$e^+e^- \to \nu \bar\nu \gamma$  with $\gamma$ emitted in the initial state
is the lone background in the $[44.5^0,135.5^0]$ angular range (white histogram).
From the same figure in this angular interval that is $-0.7<\cos\theta_\gamma <0.7$
we see that only 6 events were found, this is the real background,
not 14 events. In this case a simple method \cite{Data2010,Rick,Bayatian} is that
at 1$\sigma$ level ($68\hspace{0.5mm}\% \hspace{2mm}C.L$) for a null signal the
number of observed events should not exceed the fluctuation of the estimated
background events: $N = N_B+\sqrt{N_B}$. Of course, this method is good only when
$N_B$ is sufficiently large (i.e. when the Poisson distribution can be  approximated
with a gaussian \cite{Data2010,Rick,Bayatian}) but for $N_B > 10$ it is a good
approximation. This  means that at $1\sigma$ level ($68\hspace{0.5mm}\% \hspace{2mm}C.L.$)
the limits on the non-standard parameters are found replacing the equation for the total
number of events expected $N=6+\sqrt{6}$ in the expression $N=\sigma(\phi, \mu_{\nu_\tau}, d_{\nu_\tau}){\cal L}$.
The distributions of the photon  energy and the cosine of its polar angle are consistent
with SM predictions.

This paper is organized as follows: In Sec. II we present the
calculation of the process $e^{+}e^{-}\rightarrow \nu \bar\nu
\gamma$ in the context of a 331 model. Finally, we present our
results and conclusions in Sect. III.

\vspace{3mm}

\section{The Total Cross Section}

\vspace{3mm}

In this section we calculate the total cross section for the
reaction $e^{+}e^{-}\rightarrow \nu \bar\nu \gamma$ using the
neutral current lagrangian given in Eqs. (9) and (10) of Ref.
\cite{Cogollo} for the 331 model for diagrams 1-4 of Fig. 1. A
characteristic interesting from this model is that is independent
of the mass of the additional $Z_2$ heavy gauge boson and so we
have the mixing angle $\phi$ between the $Z_1$ and $Z_2$ bosons as
the only additional parameter. The respective transition
amplitudes are thus given by


\begin{eqnarray}
{\cal
M}_{1}&&=\frac{-g^{2}}{4\cos^{2}\theta_{W}(l^{2}-m^{2}_{\nu})}\Bigl[\bar
u(p_{3})\Gamma^{\alpha}(\l\llap{/}+m_{\nu})\gamma^{\beta}
\Bigl(\cos\phi+\frac{1-2\sin^2\theta_W}{\sqrt{3-4\sin^2\theta_W}}\sin\phi\Bigr)v(p_{4})\Bigr]\\
&&\frac{(g_{\alpha\beta}-p_{\alpha}p_{\beta}/M^{2}_{Z_1})}{\Bigl[(p_{1}+p_{2})^{2}-M^{2}_{Z_1}-i\Gamma^{2}_{Z_1}\Bigr]}\Bigl[\bar
u(p_{2})\gamma^{\alpha}\Bigl(\cos\phi-\frac{\sin\phi}{\sqrt{3-4\sin^2\theta_{W}}}\Bigr)(g_{\mbox
v}-g_{A}\gamma_{5})v(p_{1})\Bigr]\epsilon^{\lambda}_{\alpha},\nonumber
\end{eqnarray}

\begin{eqnarray}
{\cal
M}_{2}&&=\frac{-g^{2}}{4\cos^{2}\theta_{W}(l^{'2}-m^{2}_{\nu})}\Bigl[\bar
u(p_{3})\gamma^{\beta}
\Bigl(\cos\phi+\frac{1-2\sin^2\theta_W}{\sqrt{3-4\sin^2\theta_W}}\sin\phi
\Bigr)(l\llap{/}'+m_{\nu})\Gamma^{\alpha} v(p_{4})\Bigr]\\
&&\frac{(g_{\alpha\beta}-p_{\alpha}p_{\beta}/M^{2}_{Z_1})}{\Bigl[(p_{1}+p_{2})^{2}-M^{2}_{Z_1}-i\Gamma^{2}_{Z_1}\Bigr]}\Bigl[\bar
u(p_{2})\gamma^{\alpha}\Bigl(\cos\phi-\frac{\sin\phi}{\sqrt{3-4\sin^2\theta_{W}}}\Bigr)
(g_{\mbox
v}-g_{A}\gamma_{5})v(p_{1})\Bigr]\epsilon^{\lambda}_{\alpha},\nonumber
\end{eqnarray}

\begin{eqnarray}
{\cal
M}_{3}&&=\frac{-g^{2}}{4\cos^{2}\theta_{W}(k^{2}-m^{2}_{e})}\Bigl[\bar
u(p_{3})\gamma^{\alpha}\Bigl(\cos\phi+\frac{1-2\sin^2\theta_W}{\sqrt{3-4\sin^2\theta_W}}\sin\phi \Bigr)v(p_{4})\Bigr]\\
&&\frac{(g_{\alpha\beta}-p_{\alpha}p_{\beta}/M^{2}_{Z_1})}{\Bigl[(p_{1}+p_{2})^{2}-M^{2}_{Z_1}-i\Gamma^{2}_{Z_1}\Bigr]}\Bigl[\bar
u(p_{2})\gamma^{\alpha}(k\llap{/}+m_e)\gamma^{\beta}\Bigl(\cos\phi-\frac{\sin\phi}{\sqrt{3-4\sin^2\theta_{W}}}\Bigr)(g_{\mbox
v}-g_{A}\gamma_{5})v(p_{1})\Bigr]\epsilon^{\lambda}_{\alpha},\nonumber
\end{eqnarray}

\noindent and

\begin{eqnarray}
{\cal
M}_{4}&&=\frac{-g^{2}}{4\cos^{2}\theta_{W}(k^{'2}-m^{2}_{e})}\Bigl[\bar
u(p_{3})\gamma^{\alpha}\Bigl(\cos\phi+\frac{1-2\sin^2\theta_W}{\sqrt{3-4\sin^2\theta_W}}\sin\phi \Bigr)v(p_{4})\Bigr]\\
&&\frac{(g_{\alpha\beta}-p_{\alpha}p_{\beta}/M^{2}_{Z_1})}{\Bigl[(p_{1}+p_{2})^{2}-M^{2}_{Z_1}-i\Gamma^{2}_{Z_1}\Bigr]}\Bigl[\bar
u(p_{2})\gamma^{\beta}\Bigl(\cos\phi-\frac{\sin\phi}{\sqrt{3-4\sin^2\theta_{W}}}\Bigr)(g_{\mbox
v}-g_{A}\gamma_{5})(k\llap{/}'+m_e)\gamma^{\alpha}
v(p_{1})\Bigr]\epsilon^{\lambda}_{\alpha}, \nonumber
\end{eqnarray}

\noindent where

\begin{equation}
\Gamma^{\alpha}=eF_{1}(q^{2})\gamma^{\alpha}+\frac{ie}{2m_{\nu}}F_{2}(q^{2})\sigma^{\alpha
\mu}q_{\mu}+eF_3(q^2)\gamma_5\sigma^{\alpha\mu}q_\mu,
\end{equation}

\noindent

\noindent is the neutrino electromagnetic vertex, $e$ is the
charge of the electron, $q^\mu$ is the photon momentum and
$F_{1,2,3}(q^2)$ are the electromagnetic form factors of the
neutrino, corresponding to charge radius, MM and EDM,
respectively, at $q^2 = 0$ \cite{Escribano,Vogel}, while
$\epsilon^\lambda_\alpha$ is the polarization vector of the
photon. $l$ and $k$ stands for the momentum of the virtual
neutrino and antineutrino respectively.

The MM, EDM and the mixing angle $\phi$ of the 331 model give a
contribution to the total cross section for the process
$e^{+}e^{-}\rightarrow \nu \bar\nu\gamma$ of the form:

\begin{eqnarray}
\sigma(e^{+}e^{-}\rightarrow \nu\bar\nu\gamma)&=&\int
\biggl\{\frac{\alpha^2}{96\pi}\left(\kappa^2\mu^2_B+d^2_{\nu_\tau}\right)
\left[\frac{s-2\sqrt{s}E_\gamma+\frac{1}{2}E^2_\gamma\sin^2\theta_\gamma}{(s-M^2_{Z_{1}})^2+M^2_{Z_{1}}\Gamma^2_{Z_{1}}}\right]\Bigr.\nonumber\\
&+&\Bigl.\frac{\alpha^2}{64\pi}\Bigl(\kappa\mu_B+d_{\nu_\tau}\Bigr)
\Biggl(\frac{s/E^2_\gamma-2\sqrt{s}/E_\gamma}{1-\cos^2\theta_\gamma}\Biggr)\Bigr.\nonumber\\
&\times
&\Bigl.\left[\frac{\Bigl(1-s(1-2E_\gamma/\sqrt{s})/M^2_Z\Bigr)
\left((1-E_\gamma/\sqrt{s})^2+E^2_\gamma\cos^2\theta_\gamma/s\right)}
{\left(s(1-2E_\gamma/\sqrt{s})-M^2_{Z_{1}}\right)^2+M^2_{Z_{1}}\Gamma^2_{Z_{1}}}\right]\Bigr.\\
&+&\Bigl. \frac{\alpha^2}{32\pi}
\Biggl(\frac{s/E^2_\gamma-2\sqrt{s}/E_\gamma}{1-\cos^2\theta_\gamma}\Biggr)
\left[\frac{(1-E_\gamma/\sqrt{s})^2+E^2_\gamma\cos^2\theta_\gamma/s}
{\Bigl(s(1-2E_\gamma/\sqrt{s})-M^2_{Z_{1}}\Bigr)^2+M^2_{Z_{1}}\Gamma^2_{Z_{1}}}\right]\biggr\}\nonumber\\
&\times&\left[\frac{1-4x_W+8x^2_W}{x^2_W(1-x_W)^2}\right]\left(\cos\phi-\frac{\sin\phi}{\sqrt{3-4x_W}}\right)^2
\left(\cos\phi+\frac{(1-2x_W)}{\sqrt{3-4x_W}}\sin\phi\right)^2\nonumber\\
&\times &{E_\gamma dE_\gamma d\cos\theta_\gamma},\nonumber
\end{eqnarray}

\noindent where $x_{W}\equiv \sin^{2}\theta_{W}$ and $E_{\gamma}$,
$\cos\theta_{\gamma}$ are the energy and the opening angle of the
emitted photon.

It is useful to consider the smallness of the mixing angle $\phi$,
as indicated in the Eq. (14), to approximate the cross section in
Eq. (6) by its expansion in $\phi$ up to the linear term:
$\sigma=(\kappa^2\mu^2_B+d^2_{\nu_\tau})[A+B(\phi)]
+(\kappa\mu_B+d_{\nu_\tau})[C+D(\phi)]+E+F(\phi)+O(\phi^2)$, where
$A$, $B$, $C$, $D$, $E$ and $F$ are constants which can be
evaluated. Such an approximation for deriving the limits of
$\mu_{\nu_\tau}$ and $d_{\nu_\tau}$ is more illustrative and
easier to manipulate.

For $\phi< 1$, the total cross section for the process
$e^{+}e^{-}\rightarrow \nu \bar\nu\gamma$ is given by

\begin{equation}
\sigma(e^{+}e^{-}\rightarrow \nu
\bar\nu\gamma)=(\mu^2_{\nu_\tau}+d^2_{\nu_\tau})[A+B\phi]
+(\kappa\mu_B+d_{\nu_\tau})[C+D\phi]+E+F \phi+O(\phi^2),
\end{equation}

\noindent where $A$ explicitly is

\begin{equation}
A=\int \frac{\alpha^2}{96\pi}
\left[\frac{1-4x_W+8x^2_W}{x^2_W(1-x_W)^2}\right]
\left[\frac{s-2\sqrt{s}E_\gamma+\frac{1}{2}E^2_\gamma\sin^2\theta_\gamma}{(s-M^2_{Z_{1}})^2+M^2_{Z_{1}}\Gamma^2_{Z_{1}}}\right]
{E_\gamma dE_\gamma d\cos\theta_\gamma},
\end{equation}

\vspace{4mm}

\noindent while $B$, $C$, $D$, $E$ and $F$ are given by

\begin{equation}
B=\int \frac{\alpha^2}{16\pi}
\left[\frac{1-4x_W+8x^2_W}{x_W(1-x_W)^2}\right]\left[\frac{1}{\sqrt{3-4x_W}}\right]
\left[\frac{s-2\sqrt{s}E_\gamma+\frac{1}{2}E^2_\gamma\sin^2\theta_\gamma}{(s-M^2_{Z_{1}})^2+M^2_{Z_{1}}\Gamma^2_{Z_{1}}}\right]
{E_\gamma dE_\gamma d\cos\theta_\gamma},
\end{equation}


\begin{eqnarray}
C=&&\int\frac{\alpha^2}{64\pi}
\left[\frac{1-4x_W+8x^2_W}{x^2_W(1-x_W)^2}\right]
\left[\frac{s/E^2_\gamma-2\sqrt{s}/E_\gamma}{1-\cos^2\theta_\gamma}\right]\nonumber\\
&\times&\left[\frac{\Bigl(1-s(1-2E_\gamma/\sqrt{s})/M^2_Z\Bigr)
\left((1-E_\gamma/\sqrt{s})^2+E^2_\gamma\cos^2\theta_\gamma/s\right)}
{\left(s(1-2E_\gamma/\sqrt{s})-M^2_{Z_{1}}\right)^2+M^2_{Z_{1}}\Gamma^2_{Z_{1}}}\right]
{E_\gamma dE_\gamma d\cos\theta_\gamma},
\end{eqnarray}

\begin{eqnarray}
D=&&\int\frac{\alpha^2}{16\pi}
\left[\frac{1-4x_W+8x^2_W}{x_W(1-x_W)^2}\right]
\left[\frac{1}{\sqrt{3-4x_W}   }\right]
\left[\frac{s/E^2_\gamma-2\sqrt{s}/E_\gamma}{1-\cos^2\theta_\gamma}\right]\nonumber\\
&\times&\left[\frac{\Bigl(1-s(1-2E_\gamma/\sqrt{s})/M^2_Z\Bigr)
\left((1-E_\gamma/\sqrt{s})^2+E^2_\gamma\cos^2\theta_\gamma/s\right)}
{\left(s(1-2E_\gamma/\sqrt{s})-M^2_{Z_{1}}\right)^2+M^2_{Z_{1}}\Gamma^2_{Z_{1}}}\right]
{E_\gamma dE_\gamma d\cos\theta_\gamma},
\end{eqnarray}

\begin{eqnarray}
E=&&\int \frac{\alpha^2}{32\pi}
\left[\frac{1-4x_W+8x^2_W}{x_W^2(1-x_W)^2}\right]\nonumber\\
&&\times\left[\frac{s/E^2_\gamma-2\sqrt{s}/E_\gamma}{1-\cos^2\theta_\gamma}\right]
\left[\frac{(1-E_\gamma/\sqrt{s})^2+E^2_\gamma\cos^2\theta_\gamma/s}
{\Bigl(s(1-2E_\gamma/\sqrt{s})-M^2_{Z_{1}}\Bigr)^2+M^2_{Z_{1}}\Gamma^2_{Z_{1}}}\right]
{E_\gamma dE_\gamma d\cos\theta_\gamma},
\end{eqnarray}

\begin{eqnarray}
F=&&\int \frac{\alpha^2}{8\pi}
\left[\frac{1-4x_W+8x^2_W}{x_W(1-x_W)^2}\right]
\left[\frac{1}{\sqrt{3-4x_W}}\right]\nonumber\\
&&\times\left[\frac{s/E^2_\gamma-2\sqrt{s}/E_\gamma}{1-\cos^2\theta_\gamma}\right]
\left[\frac{(1-E_\gamma/\sqrt{s})^2+E^2_\gamma\cos^2\theta_\gamma/s}
{\Bigl(s(1-2E_\gamma/\sqrt{s})-M^2_{Z_{1}}\Bigr)^2+M^2_{Z_{1}}\Gamma^2_{Z_{1}}}\right]
{E_\gamma dE_\gamma d\cos\theta_\gamma}.
\end{eqnarray}

\vspace{2mm}

\noindent The expression given for A corresponds to the
cross section previously reported by T. M. Gould and I. Z.
Rothstein \cite{T.M.Gould}, while $B$, $C$, $D$, $E$ and $F$ comes
from the contribution of the 331 model, of the interference and
the SM contribution due to bremsstrahlung in which the photon is
radiated to the initial electron or positron. Evaluating the limit
when the mixing angle is $\phi=0$, the terms that depend of $\phi$
in (7) are zero and Eq. (7) is reduced to the expression (3) given
in Ref. \cite{T.M.Gould}, more the contribution of the
interference and the contribution of the SM, respectively.

\section{Results and Conclusions}

\vspace{3mm}

In order to evaluate the integral of the total cross section as a
function of the parameters of the 331 model, that is to say,
$\phi$, we require cuts on the photon angle and energy to avoid
divergences when the integral is evaluated at the important
intervals of each experiment. We integrate over $\theta_\gamma$
from $44.5^o$ to $135.5^o$ and $E_\gamma$ from 15 $GeV$ to 100
$GeV$ for various fixed values of the mixing angle
$\phi=-3.979\times 10^{-3}, 0, 1.309\times 10^{-4}$. Using the
following numerical values: $\sin^2\theta_W=0.2314$,
$M_{Z_1}=91.18$ $GeV$ and $\Gamma_{Z_1}=2.49$ $GeV$, we obtain the
cross section $\sigma=\sigma(\phi, \mu_{\nu_\tau},d_{\nu_\tau})$.

For the mixing angle $\phi$ between $Z_1$ and $Z_2$ of the 331
model, we use the reported data of Cogollo {\it et al.}
\cite{Cogollo}:

\begin{equation}
-3.979\times 10^{-3}\leq \phi \leq 1.309\times 10^{-4},
\end{equation}

\noindent with a $90 \%$ C. L. Other limits on the mixing angle
$\phi$ reported in the literature are given in Ref. \cite{Hoang}.

As was discussed in Refs. \cite{T.M.Gould,L3,Barnett,Feldman},
$N\approx\sigma(\phi, \mu_{\nu_\tau}, d_{\nu_\tau}){\cal L}$,
where $N=6+\sqrt{6}$ is the total number of $e^{+}e^{-}\rightarrow \nu\bar\nu\gamma$
events expected at $1\sigma$ level ($68 \hspace{0.5mm}\% \hspace{2mm}C.L.$)
as is mentioned in the introduction and ${\cal L}= 137$ $pb^{-1}$,
according to the data reported by the L3 Collaboration Ref. \cite{L3}
and references therein. Taking this into consideration, we can get
a limit for the tau-neutrino magnetic moment as a function of $\phi$
with $d_{\nu_\tau}=0$.

The values obtained for this limit for several values of the
$\phi$ parameter are show in Table 1.

\begin{center}
\begin{tabular}{|c|c|c|}\hline
$\phi$&$\mu_{\nu_\tau} (10^{-6}\mu_B)$&$d_{\nu_\tau}(10^{-17}e \mbox{cm})$\\
\hline \hline
$-3.979\times 10^{-3}$&[-2.57, 2.42]&[-4,95, 4.66]\\
\hline
0&[-2.60, 2.48]&[-5.01, 4.78]\\
\hline
$1.309\times 10^{-4}$&[-2.62, 2.50]&[-5.05, 4.82]\\
\hline
\end{tabular}
\end{center}

\begin{center}
Table 1. Limits on the $\mu_{\nu_\tau}$ magnetic moment and
$d_{\nu_\tau}$ electric dipole moment at $68 \%$ C. L. for
different values of the mixing angle $\phi$ \cite{Cogollo}. We
have applied the cuts used by L3 for the photon angle and energy.
\end{center}

\vspace{3mm}

The results obtained in Table 1 are in agreement with the
literature
\cite{T.M.Gould,H.Grotch,L3,Escribano,DELPHI,J.M.Hernandez,Perez,F.Larios,Gutierrez1,Gutierrez2,Gutierrez3,Gutierrez4,Gutierrez5,Aydin}.
However, if the mixing angle is $\phi=-2.1\times 10^{-3}, 0,
1.32\times 10^{-4}$ \cite{Hoang}, we obtained the results given in Table 2.\\

\begin{center}
\begin{tabular}{|c|c|c|}\hline
$\phi$&$\mu_{\nu_\tau} (10^{-6}\mu_B)$&$d_{\nu_\tau}(10^{-17}e \mbox{cm})$\\
\hline \hline
$-2.1\times 10^{-3}$&[-2.59, 2.44]&[-4.99, 4.70]\\
\hline
0&[-2.60, 2.48]&[-5.01, 4.78]\\
\hline
$1.32\times 10^{-4}$&[-2.64, 2.52]&[-5.09, 4.86]\\
\hline
\end{tabular}
\end{center}

\begin{center}
Table 2. Limits on the $\mu_{\nu_\tau}$ magnetic moment and
$d_{\nu_\tau}$ electric dipole moment at $68 \%$ C. L. for
different values of the mixing angle $\phi$ \cite{Hoang}. We have
applied the cuts used by L3 for the photon angle and energy.
\end{center}

\vspace*{3mm}

The previous analysis and comments can readily be translated to
the EDM of the $\tau$-neutrino with $\mu_{\nu_\tau}=0$. The
resulting limits for the EDM as a function of $\phi$ are shown in
Tables 1 and 2.

The incorporation of the diagrams with photon radiation in the initial
state, as well as the statistical analysis gives a contribution of
about $22\hspace{0.5mm}\%$  on the bounds of magnetic and electric
dipole moments of the tau-neutrino, with respect to analysis in
$Z_1$ boson resonance, that is to say $s = M^2_{Z_1}$.

We plot the total cross section in Fig. 2 as a function of the
mixing angle $\phi$ for the limits of the magnetic moment given in
Tables 1 and 2. Our results for the dependence of the differential
cross section on the photon energy versus the cosine of the
opening angle between the photon and the beam direction
$(\theta_\gamma)$ are presented in Fig. 3 for $\phi=-3.979\times
10^{-3}$ and $\mu_{\nu_\tau}=2.42\times 10^{-6}\mu_B$. In
addition, the form of the distributions does not change
significantly for the values $\phi$ and $\mu_{\nu_\tau}$ because
$\phi$ and $\mu_{\nu_\tau}$ are very small in value, as shown in
Tables 1-2. Finally, we plot the differential cross-section in
Fig. 4 as a function of the photon energy for the limits of the
magnetic moments given in Tables 1-2.

Other upper limits on the tau-neutrino magnetic moment reported in
the literature are $\mu_{\nu_{\tau}} < 3.3 \times 10^{-6} \mu_{B}$
$(90 \hspace{1mm}\% \hspace{1mm} C.L.)$ from a sample of
$e^{+}e^{-}$ annihilation events collected with the L3 detector at
the $Z_1$ resonance corresponding to an integrated luminosity of
$137$ $pb^{-1}$ \cite{L3}; $\mu_{\nu_{\tau}} \leq 2.7 \times
10^{-6} \mu_{B}$ $(95 \hspace{1mm} \% \hspace{1mm} C.L.)$ at
$q^2=M^2_{Z_1}$ from measurements of the $Z_1$ invisible width at
LEP \cite{Escribano}; $\mu_{\nu_{\tau}} \leq 2.62 \times 10^{-6}$
in the effective Lagrangian approach at the $Z_1$ pole
\cite{Maya}; $\mu_{\nu_{\tau}} < 1.83 \times 10^{-6} \mu_{B}$ $(90
\hspace{1mm} \% \hspace{1mm}C.L.)$ from the analysis of
$e^{+}e^{-}\rightarrow \nu \bar\nu \gamma$ at the $Z_1$-pole, in a
class of $E_6$ inspired models with a light additional neutral
vector boson \cite{Aytekin}; from the order of $\mu_{\nu_{\tau}} <
O(1.1 \times 10^{-6} \mu_{B})$ Keiichi Akama {\it et al.} derive
and apply model-independent limits on the anomalous magnetic
moments and the electric dipole moments of leptons and quarks due
to new physics \cite{Keiichi}. However, the limits obtained in
Ref. \cite{Keiichi} are for the tau-neutrino with an upper bound
of $m_\tau < 18.2$ $MeV$ which is the current experimental limit.
It was pointed out in Ref. \cite{Keiichi} however, that the upper
limit on the mass of the electron neutrino and data from various
neutrino oscillation experiments together imply that none of the
active neutrino mass eigenstates is heavier than approximately 3
$eV$. In this case, the limits given in Ref. \cite{Keiichi} are
improved by seven orders of magnitude. The limit $\mu_{\nu_{\tau}}
< 5.4 \times 10^{-7} \mu_{B}$ $(90 \hspace{1mm} \% \hspace{1mm}
C.L.)$ is obtained at $q^2=0$ from a beam-dump experiment with
assumptions on the $D_s$ production cross section and its
branching ratio into $\tau \nu_\tau$ \cite{A.M.Cooper}, thus
severely restricting the cosmological annihilation scenario
\cite{G.F.Giudice}. Our results in Tables 1 and 2 for
$\phi=-3.979\times 10^{-3}, 0, 1.309\times 10^{-4}$ and
$\phi=-2.1\times 10^{-3}, 0, 1.32\times 10^{-4}$ compare favorably
with the limits obtained by the L3 Collaboration \cite{L3}, and
with others limits reported in the literature \cite{T.M.Gould,H.Grotch,Escribano,Maya}.

In the case of the electric dipole moment, other upper limits
reported in the literature are: $\mid d(\nu_{\tau})\mid \leq 5.2
\times 10^{-17} \mbox{$e$ cm}, \hspace*{1mm} {\mbox {95\% C.L.}}$
\cite{Escribano} and $\mid d(\nu_{\tau})\mid < O(2 \times 10^{-17}
\mbox{$e$ cm})$ \cite{Keiichi}.

In summary, we conclude that the estimated limits for the tau-
neutrino magnetic and electric dipole moments in the context of a
331 model compare favorably with the limits obtained by the L3
Collaboration, and complement previous studies on the dipole
moments. In the limit $\phi=0$ our limits takes the value
previously reported in Ref. \cite{T.M.Gould} for the SM. On the
other hand, it seems that in order to improve these limits it
might be necessary to study direct CP-violating effects
\cite{M.A.Perez}. In addition, the analytical and numerical
results for the total cross section have never been reported in
the literature before and could be of some practical use for the
scientific community.

\vspace{8mm}

\begin{center}
{\bf Acknowledgments}
\end{center}

We acknowledge support from CONACyT, SNI and PROMEP (M\'exico).

\newpage

\begin{figure}[t]
\centerline{\scalebox{0.6}{\includegraphics{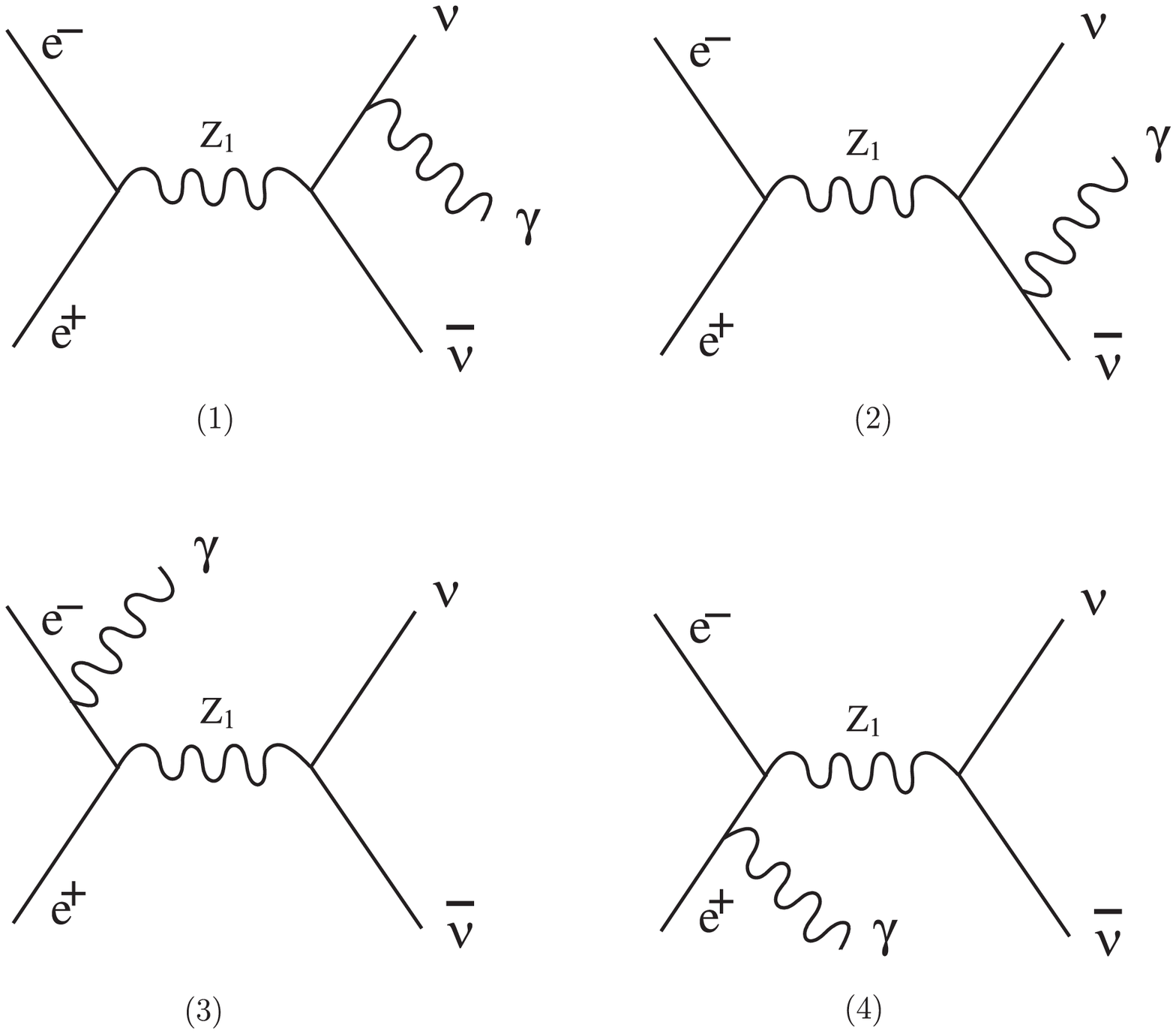}}}
\caption{ \label{fig:gamma} The Feynman diagrams contributing to
the process $e^{+}e^{-}\rightarrow \nu\bar\nu\gamma$ in a 331
model (1, 2) when the $Z_1$ vector boson is produced on mass-shell
and the SM (3, 4) diagrams for initial-state radiation.}
\end{figure}

\begin{figure}[t]
\centerline{\scalebox{1.2}{\includegraphics{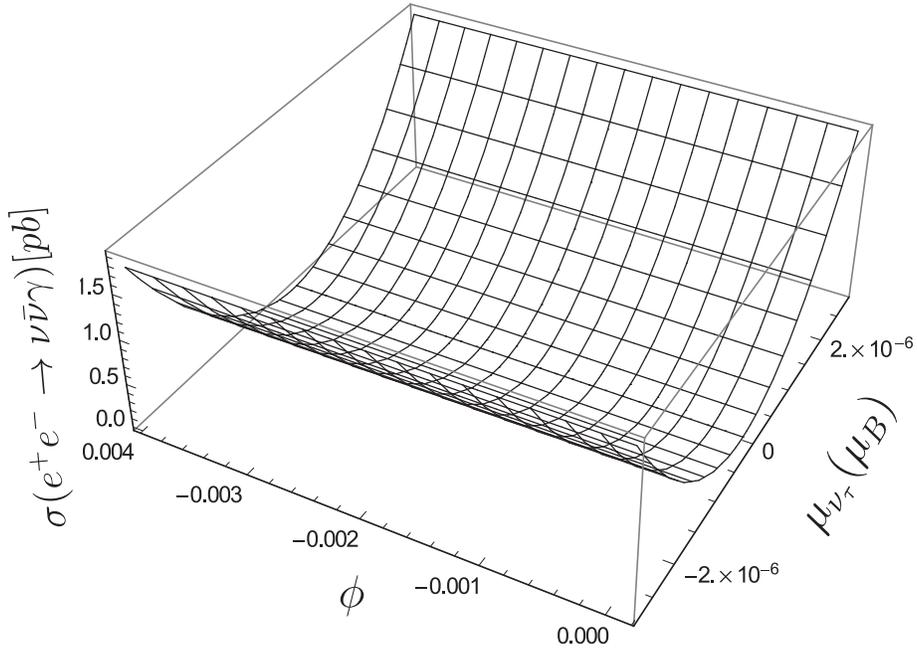}}}
\caption{ \label{fig:gamma} The total cross section for
$e^{+}e^{-}\rightarrow \nu \bar\nu\gamma$ as a function of $\phi$
and $\mu_{\nu_\tau}$ (Tables 1, 2).}
\end{figure}

\begin{figure}[t]
\centerline{\scalebox{1.2}{\includegraphics{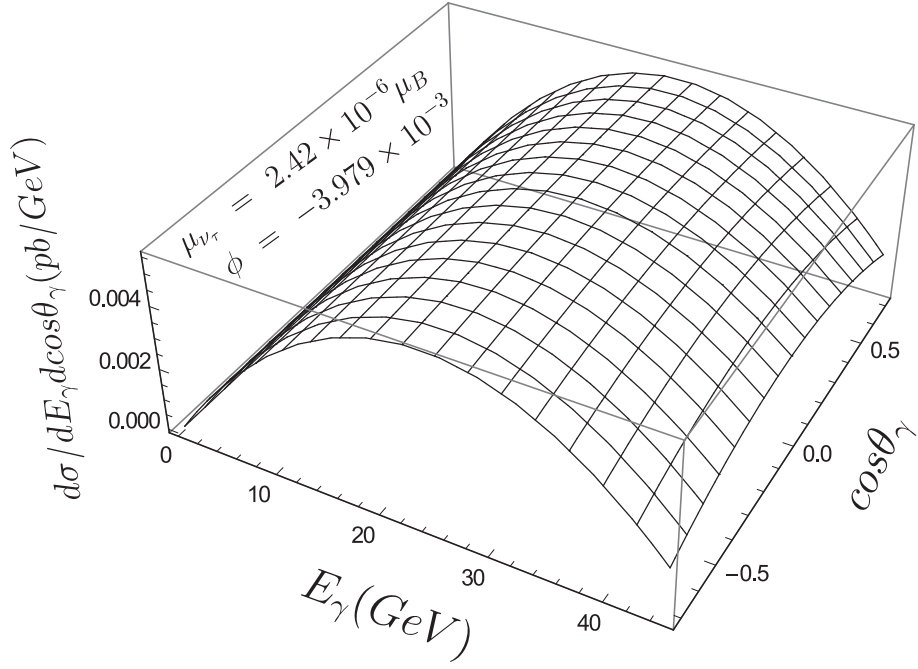}}}
\caption{ \label{fig:gamma} The differential cross section for
$e^{+}e^{-}\rightarrow \nu\bar\nu\gamma$ as a function of
$E_\gamma$ and $\cos\theta_\gamma$ for $\phi=-3.979\times 10^{-3}$
and $\mu_{\nu_\tau}=2.42\times 10^{-6}\mu_B$.}
\end{figure}

\begin{figure}[t]
\centerline{\scalebox{1.2}{\includegraphics{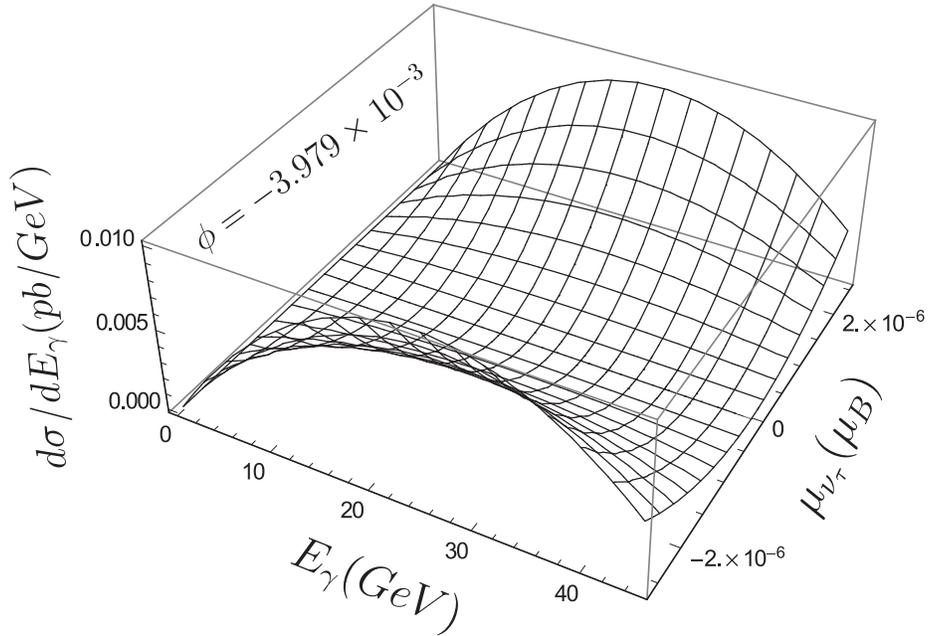}}}
\caption{ \label{fig:gamma} The differential cross section for
$e^{+}e^{-}\rightarrow \nu\bar\nu\gamma$ as a function of
$E_\gamma$ and $\mu_{\nu_\tau}$ with $\phi=-3.979\times 10^{-3}$.}
\end{figure}

\end{document}